\documentclass[a4paper,12pt]{article}
\usepackage{amsmath}
\usepackage{amssymb}
\usepackage{amsthm}
\usepackage{latexsym}
\usepackage{mathrsfs}
\usepackage[dvips]{graphicx}
\usepackage{pstricks}
\usepackage{color}
\def\f(#1){{\mathop{f}^{(#1)}}}
\def\m(#1){{\mathop{m}^{(#1)}}}
\def\C(#1){{\mathop{C}^{(#1)}}}
\def\p(#1){{\mathop{p}^{(#1)}}}

\def\ben{\begin{equation}}
\def\een{\end{equation}}
\def\bena{\begin{eqnarray}}
\def\eena{\end{eqnarray}}
\def\non{\nonumber}

\def\d{{\rm d}}

\def\C{{\cal C}}

\def\mr{{\mathbb R}}

\newcommand{\mz}{{\mathbb Z}}

\newcommand{\e}{{\rm e}}

\theoremstyle{theorem}

\newtheorem{thm}{Theorem}

\begin{document}

\title{All Vacuum Near-Horizon Geometries in $D$-dimensions with 
$(D-3)$ Commuting Rotational Symmetries}

\author{
Stefan Hollands$^{1,2}$\thanks{\tt HollandsS@Cardiff.ac.uk}\: and
Akihiro Ishibashi$^{2}$\thanks{\tt akihiro.ishibashi@kek.jp}\:
\\ \\
{\it ${}^{1}$School of Mathematics, Cardiff University} \\
{\it Cardiff, United Kingdom} \medskip \\
{\it ${}^{2}$KEK Theory Center,
Institute of Particle and Nuclear Studies
} \\
{\it High Energy Accelerator Research Organization (KEK)
} \\
{\it Tsukuba, Japan} \\
}

\date{19 January 2010}
\maketitle

\begin{abstract}
We explicitly construct all stationary, non-static, extremal near horizon 
geometries in $D$ dimensions that satisfy the vacuum Einstein equations, 
and that have $D-3$ commuting rotational symmetries. 
Our work generalizes [arXiv:0806.2051] by Kunduri and Lucietti, where such 
a classification had been given in $D=4,5$. But our method is different from 
theirs and relies on a matrix formulation of the Einstein equations. 
Unlike their method, this matrix formulation works for any dimension. 
The metrics that we find come in three families, with horizon topology $S^2 \times T^{D-4}$, or
$S^3 \times T^{D-5}$, or quotients thereof. Our metrics depend on two
discrete parameters specifying the topology type, as well as 
$(D-2)(D-3)/2$ continuous parameters.  
Not all of our metrics in $D \ge 6$ seem to arise as the near horizon limits of known
black hole solutions.
\end{abstract}


\sloppy

\section{Introduction}  

Many known families of black hole solutions possess a limit wherein the black hole horizon becomes degenerate,
i.e. where the surface gravity tends to zero; such black holes are called extremal. While extremal black holes
are not believed to be physically realized as macroscopic objects in nature, they are nevertheless highly interesting
from the theoretical viewpoint. Due to the limiting procedure, they are in some sense at the fringe of
the space of all black holes, and therefore possess special properties which make them easier to study
in various respects. For example, in string theory, the derivation of the Bekenstein-Hawking entropy of black holes from counting
microstates (see e.g.~\cite{david} for a review)
is best understood for extremal black holes. Furthermore, many black hole solutions
that have been constructed in the context of supergravity theories (see e.g.~\cite{Gauntlett1,Gauntlett2})
have supersymmetries, and are thus automatically extremal.

Many of the arguments related to the derivation of the black hole entropy---especially in the context of the
``Kerr-CFT correspondence''~\cite{guida,lu,compere,aze,Hartman,amsel}---actually only involve the spacetime geometry
in the immediate (actually infinitesimal) neighborhood of the black hole horizon. More precisely, by applying a suitable scaling process to the spacetime metric which in effect blows up this neighborhood, one can obtain in the limit a new spacetime metric, called a ``near horizon geometry.'' It is the near horizon geometry which enters many of the arguments pertaining to
the derivation of the black hole entropy.

The near horizon limit can be defined for any spacetime $(M,g)$ with a degenerate Killing horizon, $N$---not necessarily a black hole horizon. The construction runs as follows\footnote{ 
The general definition of a near-horizon limit was first considered 
in the context of supergravity black holes in \cite{Reall03},  
and in the context of extremal but not supersymmetric black holes 
in \cite{crt06} for the static case and in \cite{klr} for the general case. 
The concept of near-horizon geometry itself has appeared previously 
in the literature, e.g., \cite{Hajicek} for $4$-dimensional vacuum case (also see \cite{lp03} for the isolated horizon case).  
} 
. 
First, recall that a spacetime with degenerate Killing horizon by definition has a smooth, codimension one, null hypersurface $N$, and a Killing vector field $K$ whose orbits are tangent $N$, and which on $N$ are tangent to affinely\footnote{For a non-degenerate horizon, the orbits on $N$ of $K$ would not be affinely
parametrized.} parametrized null-geodesics.
Furthermore, by assumption, there is a ``cross section'', $H$, of codimension one in $N$ with the property
that each generator of $K$ on $N$ is isomorphic to $\mr$ and intersects $H$ precisely once.
In the vicinity of $N$, one can then introduce
``Gaussian null coordinates'' $u,v,y^a$ as follows, 
see e.g. \cite{MI83}.  
First, we choose arbitrarily
local\footnote{Of course, it will take more than one patch to cover $H$, but
the fields $\gamma, \beta, \alpha$ on $H$ below in eq.~\eqref{gnc} are globally defined and
independent of the choice of coordinate systems.} coordinates $y^a$ on $H$, and we Lie-transport them along the flow of $K$ to other places on $N$,
denoting by $v$ the flow parameter. Then, at each point of $N$ we shoot off affinely parametrized null-geodesics and take $u$ to be the affine parameter along these null geodesics. The tangent vector $\partial/\partial u$
to these null geodesics is required to have unit inner product with $K = \partial/\partial v$ on $H$, and to be orthogonal
to the Lie-transported cross-section $H$. It can be shown that the metric then takes the Gaussian null form
\ben\label{gnc}
g = 2 \d v (\d u  + u^2 \alpha \, \d v + u \beta_a \, \d y^a) + \gamma_{ab} \, \d y^a \d y^b \, ,
\een
where the function $\alpha$, the one-form $\beta = \beta_a \, \d y^a$, and the tensor field
$\gamma = \gamma_{ab} \, \d y^a \d y^b$ do not depend on $v$. The Killing horizon $N$ is
located at $u=0$, and the cross section $H$ at $u=v= 0$. The near horizon limit is now taken 
by applying to $g$ the diffeomorphism $v \mapsto v/\epsilon, u \mapsto \epsilon u$ (leaving the other coordinates $y^a$ unchanged), and then taking $\epsilon \to 0$. The so-obtained metric looks
exactly like eq.~\eqref{gnc}, but with new metric functions obtained from the old ones by evaluating them at $u=0$. Thus, the fields $\alpha, \beta, \gamma$ of the near horizon metric neither depend on $v$ nor $u$, and
can therefore be viewed as fields on $H$. If the original spacetime with degenerate Killing horizon satisfied the
vacuum Einstein equation or the Einstein equation with a cosmological constant, then the near horizon limit does, too.

The near horizon limit is simpler than the original metric in the sense that it has more symmetries. For example, if the limit procedure is applied to the extremal Kerr metric in $D=4$ spacetime dimensions with symmetry group
$\mr \times U(1)$, then---as observed\footnote{By construction, the near horizon geometry has the Killing fields
$\partial/\partial  v$ and $u \partial/\partial u - v \partial/\partial v$, which
generate a two-parameter symmetry group. The non-trivial observation by~\cite{bardeen} is that this actually
gets enhanced to the three-parameter group $O(2,1)$.}
first by~\cite{bardeen} (see also \cite{BW,Carter})---the near horizon 
metric has an enhanced symmetry
group of $O(2,1) \times U(1)$. The first factor of this group is related to an $AdS_2$-factor in the metric. A similar phenomenon occurs for stationary extremal black holes in higher dimensions with a
comparable amount of symmetry: As proved in~\cite{klr}, if $(M,g)$ is a $D$-dimensional
stationary extremal black hole with isometry group\footnote{The ``rigidity theorem''~\cite{hi}
guarantees that a stationary extremal black hole has a symmetry group that contains $\mr \times U(1)$,
i.e. guarantees only one axial Killing field in addition to the assumed timelike Killing field. Therefore,
in $D \ge 5$, assuming a factor of $U(1)^{D-3}$ is a non-trivial restriction, while it is actually a
consequence of the rigidity theorem in $D=4$.} $\mr \times U(1)^{D-3}$ and compact horizon cross section $H$,
then the near horizon limit has the enhanced symmetry group $O(2,1) \times U(1)^{D-3}$.
In $D \ge 5$ dimensions, it is not known at present what is the most 
general stationary extremal black hole solution with symmetry group 
$\mr \times U(1)^{D-3}$, so one can neither perform explicitly 
their near horizon limits. 
Nevertheless, because
the near horizon metric has an even higher degree of symmetry---the metric functions essentially only depend non-trivially
on one coordinate---one can try to classify them directly.

This was done for the vacuum Einstein equations
in dimensions $D=4,5$ by~\cite{kl}, where a list of all near horizon geometries, i.e. metrics of the form~\eqref{gnc} with metric functions $\alpha, \beta, \gamma$ independent of $u,v$, was obtained. It is a priori
far from obvious that {\em all} these metrics are the near horizon limits of actual globally defined
black holes. Remarkably though,~\cite{kl} could prove that the metrics found are indeed the limits of the extremal black ring~\cite{Emparan}, boosted Kerr string, Myers-Perry~\cite{Myers}, and the Kaluza-Klein black holes~\cite{ras,lar}, respectively.

In this paper, we give a classification of all possible vacuum near horizon geometries with
symmetry group $O(2,1) \times U(1)^{D-3}$ in arbitrary dimensions $D$. 
The method of analysis used in~\cite{kl} seems restricted to $D=4,5$, so we here use a different method based on a matrix formulation of the vacuum Einstein equations that works in arbitrary dimensions. The metrics that we find come in three families
depending on the topology of $H$, which can be either $S^3 \times T^{D-5}, S^2 \times T^{D-4}$ or $L(p,q) \times T^{D-5}$,
where $L(p,q)$ is a Lens space.
The metrics in each of these families depend on 
$(D-2)(D-3)/2$ real parameters;  
they are given explicitly in Thm.~1 below. 
When specialized to $D=5$, our 
first two families of metrics must coincide with those previously found in~\cite{kl}, whereas the last family
is shown to arise from the first one by taking quotients (this last properties generalizes to
arbitrary $D$). 
In all dimensions, examples for 
near horizon geometries with topology $S^2\times T^{D-4}$ are 
provided by the near horizon limit of the ``boosted Kerr-branes'' 
see e.g. \cite{klr,fklr08}. 
This family of metrics depends on $(D-2)(D-3)/2$ real parameters 
and it is conceivable that all near horizon geometries of 
this topology can be obtained in this way.  
The analogous construction is also possible 
when the horizon topology is $S^3 \times T^{D-5}$. 
However, in this case, the resulting metrics depend on fewer 
parameters. 
%
%
%

We should also point out that there are vacuum near-horizon 
geometries that possess fewer symmetries than $\mr \times U(1)^{D-3}$. 
For example, the near-horizon geometry of the extremal Myers-Perry 
black holes, constructed explicitly 
in \cite{fklr08}, has the smaller symmetry group, 
$\mr \times U(1)^{[(D-1)/2]}$.  
In this paper we are not going to classify such less symmetric 
vacuum near-horizon geometries. 
Also, we are not going to consider the case of a non-vanishing 
cosmological constant, since, as far as we are aware, there has 
appeared no successful reduction of the Einstein gravity with 
a cosmological constant to a suitable nonlinear sigma model, which is 
however required in our approach.  
The same remark would apply to other theories with different 
matter fields. On the other hand, we expect our approach to be 
applicable to theories that can be reduced to suitable sigma-models. 
For $D=5$ minimal gauged and ungauged supergravity, the near horizon 
geometries were classified in \cite{klr2,Reall03} using a method different 
from ours. Also for $D=4$ Einstein-Maxwell theory with a cosmological 
constant, see e.g. \cite{kl2}.

%

\section{Geometrical coordinates}

The aim of this paper is to classify the near horizon geometries in $D$ dimensions. As explained in the
previous section, by this we mean the problem of finding all metrics $g$ of the form~\eqref{gnc} with vanishing Ricci tensor
(i.e. vacuum metrics), where $\gamma = \gamma_{ab} \d y^a \d y^b$ is a smooth metric on the compact
manifold $H$, $\beta = \beta_a \d y^a$ is a 1-form on $H$ and $\alpha$ is a scalar function on $H$.
These fields do not depend on $u,v$, and the near horizon geometries therefore
have the Killing vectors $K = \partial/\partial v$ and $X = u \partial/\partial u - v \partial/\partial v$. We do {\em not} assume a priori that the near horizon metrics arise from
a black hole spacetime by the limiting procedure described above.

Unfortunately, this problem appears to be difficult to solve in this generality, so we will make a significant further symmetry assumption. Namely, we will assume that our metrics do not only have
the Killing vectors $K,X$, but in addition admit the symmetry group $U(1)^{D-3}$, generated by
$(D-3)$ commuting Killing fields $\psi_1, \dots, \psi_{D-3}$ that are tangent to $H$ and also commute with $K,X$.
Thus, the full isometry group of our metric is (at least) $G_2 \times U(1)^{D-3}$, where $G_2$ denotes the Lie-group that
is generated by $K, X$. This means roughly speaking that the metric
functions can nontrivially depend only on a single variable, and our metrics may
hence be called ``cohomogeneity-one.'' As a consequence,
Einstein's equations reduce to a coupled system of non-linear ordinary differential equations in
this variable. Our aim is to solve this system in the most general way and thereby to classify all near
horizon geometries with the assumed symmetry.

It seems that this system becomes tractable only if certain special coordinates are introduced that are adapted in an
optimal way to the geometric situation under consideration. These coordinates are the well-known Weyl-Papapetrou
coordinates up to a simple coordinate transformation. However, to introduce these coordinates
in a rigorous and careful manner is more subtle in the present case than for non-extremal horizons.
These technical difficulties are closely related to the fact that the usual Weyl-Papapetrou coordinates are actually singular on $H$, the very place we are interested in most. To circumvent this problem, we follow the elegant alternative procedure introduced in~\cite{klr,kl}. That procedure applies in the form presented here to non-static geometries, and we will
for the rest of this paper make this assumption.
The static case has been treated previously in \cite{crt06,kl1}. 

We first observe that the horizon $H$ is a compact $(D-2)$-dimensional manifold with an action of $U(1)^{D-3}$.
By general and rather straightforward arguments (see e.g.~\cite{pak,hs}) it follows that, topologically, $H$ can only be of the following four types:
\ben
H \cong \begin{cases}
S^3 \times T^{D-5} \, ,\\
S^2 \times T^{D-4} \, ,\\
L(p,q) \times T^{D-5} \, ,\\
T^{D-2} \, .
\end{cases}
\een
Furthermore, in the first three cases, the quotient space $H/U(1)^{D-3}$ is a closed interval---which we
take to be $[-1,1]$ for definiteness---whereas in
the last case, it is $S^1$. We will not treat the last case 
in this paper\footnote{ 
See, however, the note added in proof. 
}
, but we note that
the topological censorship theorem~\cite{galoway} implies that there cannot exist any extremal, asymptotically flat or Kaluza-Klein vacuum black holes with $H \cong T^{D-2}$. Thus, while there could still be near
horizon geometries with $H \cong T^{D-2}$, they cannot arise as the limit of a globally defined black hole spacetime.

In this paper, we will focus on the first three topology types. In these cases, the Gram matrix
\ben\label{gram}
f_{ij} = \gamma(\psi_i, \psi_j)
\een
is non-singular in the interior of the interval and it has a one-dimensional null-space
at each of the two end points~\cite{hs}. In fact, there are integers $a_\pm^i \in \mz$ such that
\ben\label{bndycond}
f_{ij}(x) a_\pm^i \to 0 \quad \text{at boundary points $\pm 1$.}
\een
The integers $a^i_\pm$ determine the topology of $H$ (i.e. which of the first three cases
we are in), as we explain more in Thm.~\ref{thm1} below.

The first geometric coordinate, $x$, parametrizes the interval $[-1,+1]$, and is introduced
as follows. Consider the 1-form on $H$ defined by $\Sigma = (\det f)
\star_\gamma (\psi_1 \wedge \dots \wedge \psi_{D-3})$, where the Hodge
dual is taken with respect to the metric $\gamma$ on $H$. Using the fact that
the $\psi_i$ are commuting Killing fields of $\gamma$, one can show that $\Sigma$ is closed, and
that it is Lie-derived by all $\psi_i$. Hence $\Sigma$ may be viewed as a closed 1-form on the orbit
space $H/U(1)^{D-3}$, which, as we have said, is a closed interval. It can be seen furthermore that $\Sigma$ does not
vanish anywhere within this closed interval, so there exists a function $x$, such that
\ben
\d x = C \Sigma \, .
\een
The constant $C$ is chosen so that $x$ runs from $-1$ to $+1$. We take $x$ to be our first coordinate, and we
take the remaining coordinates on $H$ to be angles $\varphi^1, \dots, \varphi^{D-3}$
running between $0$ and $2\pi$, chosen in such a way
that $\psi_i = \partial/\partial \varphi^i$. In these coordinates, the metric $\gamma$ on $H$ takes the form
\ben
\gamma = \frac{1}{C^2 \det f} \, \d x^2 + f_{ij}(x) \d \varphi^i \d \varphi^j \, .
\een
To define our next coordinate, we consider the 1-form field $\beta$ on $H$, see eq.~\eqref{gnc}.
Standard results on the Laplace operator $\Delta_\gamma$ on a compact Riemannian manifold $(H, \gamma)$
guarantee that there exists a smooth function $\lambda$ on $H$ such that
\ben\label{lamdef}
\star_\gamma \d\! \star_\gamma \beta = \Delta_\gamma \lambda \, ,
\een
where $\star_\gamma$ is the Hodge star of $\gamma$. The function $\lambda$ is unique up to a constant. Because
$\beta$ and $\gamma$ are Lie-derived by all the rotational Killing fields $\psi_i$, it follows that
${\mathscr L}_{\psi_i} \lambda = c_i$ are harmonic functions on $H$, i.e. constants. Furthermore, these constants
must vanish, because the $\psi_i$ have periodic orbits. Thus, $\lambda$ is only a function of $x$.
We also claim that the 1-form $\beta - \d \lambda$ has no $\d x$-part. To see this, we let
$h$ be the scalar function on $H$ defined by $h = \star_\gamma (\psi_1 \wedge \dots \wedge \psi_{D-3} \wedge [\beta - \d \lambda])$.
Using eq.~\eqref{lamdef} and the fact that the $\psi_i$ are commuting Killing fields of $\gamma$, it is easy to
show that $\d h = 0$, so $h$ is constant. Furthermore, by eq.~\eqref{bndycond} there exist points in $H$ where the
linear combinations $a^i_\pm \psi_i = 0$, and it immediately follows from this that $h = 0$ on $H$. This shows that
$\beta - \d \lambda$ has no $\d x$-part, hence we can write
\ben
\beta = \d \lambda + C \e^\lambda \, k_i \d\varphi^i \, ,
\een
where we have introduced the quantities
\ben
k_i := C^{-1} \e^{-\lambda} \, \psi_i \cdot \beta \, .
\een
The next coordinate is defined by
\ben
r:= u \e^\lambda \, ,
\een
and we keep $v$ as the last remaining coordinate. The coordinates
$\varphi^i, r, x, v$ are the desired geometrical coordinates. In these,
the metric takes the form
\ben
g = \e^{-\lambda} [2 \d v \d r + r^2 (2\alpha \e^{-\lambda} - \e^\lambda k_i k^i) \, \d v^2]
+ \frac{\d x^2}{C^2 \det f} + f_{ij}(\d \varphi^i + Cr \, k^i \d v)(\d \varphi^j + Cr \, k^j \d v) \, .
\een
We have also determined that the quantities $k^i, f_{ij}, \alpha, \lambda$ are functions of $x$ only.
The indices $i,j,...$ are raised with the inverse $f^{ij}$ of the Gram matrix, e.g. $k^i = f^{ij} k_j$.

So far, we have only used the symmetries of the metric, but not the fact that it is also required to be Ricci flat.
This imposes significant further restrictions~\cite{kl,klr}. 
Namely, one finds that $k^i$ are simply {\em constants},
and that $(2\alpha \e^{-\lambda} - \e^\lambda k_i k^i)$
is a negative\footnote{Here one must use that the metric is {\em not} static, i.e. that not all
 $k^i$ vanish.} {\em constant}, which one may choose to be $-C^2$ after a suitable rescaling of the
coordinates $r,v$ and the constants $k^i$, and by adding a constant to $\lambda$. Then the Einstein
equations further imply that $\partial_x^2(\e^{-\lambda} \det f)=-2$; hence $\e^{-\lambda} = -(x-x_-)(x-x_+)(\det f)^{-1}$
for real numbers $x_\pm$. Furthermore, $\lambda$ is smooth and $\det f$ vanishes only at
$x = \pm 1$ by eq.~\eqref{bndycond}, so $x_\pm = \pm 1$ and consequently
\ben\label{lambdax}
\e^{-\lambda} = (1-x^2)(\det f)^{-1} \, .
\een
Thus, in summary, we have determined that the near horizon metric
is given by
\ben\label{nhg}
g = \frac{1-x^2}{\det f}(2\d v \d r - C^2 r^2 \d v^2) + \frac{\d x^2}{C^2 \det f}  +
f_{ij}( \d \varphi^i + r Ck^i \, \d v)(\d \varphi^j + r Ck^j \, \d v)
\een
where $k^i, C$ are constants, and where $f_{ij}$ depends only on $x$.

In the remainder of the paper, we will work with above form of the metric~\eqref{nhg}. However, we will,
for completeness, also give the relation to the more familiar Weyl-Papapetrou form: For
$r > 0$ (i.e., strictly outside the horizon), we define new coordinates $(t,\rho,z,\phi^i)$ by the transformation \cite{fl} 
\bena\label{rx}
z &:=& rx\\
\rho &:=& r\sqrt{1-x^2}\\
t &:=& Cv + (Cr)^{-1}\\
\phi^i &:=& \varphi^i + C^{-1} k^i \log r \, .
\eena
In the new coordinates $(t,\rho,z,\phi^i)$, the metric then takes the Weyl-Papapetrou form
\ben\label{wp}
g = - \frac{\rho^2 \, \d t^2}{\det f} + \frac{\e^{-\lambda}}{C^2r^2} (\d \rho^2 + \d z^2)
+
f_{ij}( \d \phi^i + r k^i \, \d t)(\d \phi^j + r k^j \, \d t) \, ,
\een
where it is understood that $r^2 = \rho^2 + z^2$.
Note that, by contrast with the coordinate system $(v,r,x,\varphi^i)$, the Weyl-Papapetrou coordinate system
does not cover the horizon itself, i.e., it is not defined for $r=0$ but only for $r>0$. This can be seen
in several ways, for example by noting that the coordinate transformation is singular at $r=0$, i.e.
on the horizon, or alternatively,
by noting that the horizon corresponds in the new coordinates to the single point $\rho = z = 0$. This
behavior is characteristic for extremal horizons and does not happen in the non-extremal case.

In obtaining our form~\eqref{nhg} for the near horizon metric, we have used up all but the $ij$-components of the
Einstein equations. The remaining Einstein equations determine the matrix of functions
$f_{ij}(x)$. As is well-known~\cite{Maison}, a beautifully simple form of these equations
can be obtained by introducing the twist potentials
of the rotational Killing fields as auxiliary variables. These potentials $\chi_i$ are defined up to a constant by
\ben
\d \chi_i = \star ( \psi_1 \wedge \dots \wedge \psi_{D-3} \wedge \d \psi_i) \, .
\een
To see that this equation makes sense, one has to prove that the right side is an exact form. Indeed, taking
$\d$ of the right side and using the vanishing of the Ricci tensor together with the fact that the Killing fields all commute,
one gets zero. To see that the right side is even exact, it is best to pass to the orbit space $M/(G_2 \times U(1)^{D-3})$ first,
which can be identified with the interval $[-1,1]$. Then the $\chi_i$ can be defined on this orbit space and lifted back to functions
on $M$. It also follows from this construction that $\chi_i$ only depends
on the coordinate $x$ parametrizing $[-1,1]$. Setting
\ben\label{phidef}
\Phi = \left(
\begin{matrix}
(\det f)^{-1} & -(\det f)^{-1} \chi_i \\
-(\det f)^{-1} \chi_i & f_{ij} + (\det f)^{-1} \chi_i \chi_j
\end{matrix}
\right) \quad ,
\een
it is well-known that the vanishing of the Ricci-tensor implies that
\ben\label{phirx}
\partial_x [(1-x^2) \Phi^{-1} \partial_x \Phi] + \partial_r[ r^2 \, \Phi^{-1} \partial_r \Phi] = 0 \, .
\een
These equations are normally written in the Weyl-Papapetrou coordinates $\rho,z$ (see e.g.~\cite{hs}), and the above form is obtained simply by the change of variables eq.~\eqref{rx}.

Since $\Phi$ is a function of $x$ only in our situation (but would not be e.g.
for black holes without the near horizon limit taken) an essential further simplification occurs: The second term in
the above set of matrix equations is simply zero! Hence, the content of the remaining Einstein equations is
expressed in the matrix of {\em ordinary} differential equations
\ben\label{phieq}
\partial_x [(1-x^2) \Phi^{-1} \partial_x \Phi] = 0 \, .
\een
In fact, this equation could be derived formally and much more directly by simply assuming the Weyl-Papapetrou
form of the metric, introducing $r,x$ as above, and then observing that, in the near horizon limit,
the dependence on $r$ is scaled away, so that the matrix partial differential equations~\eqref{phirx} reduce to the ordinary
differential equations~\eqref{phieq}.

\section{Classification}

To determine all near horizon metrics~\eqref{nhg}, we must solve the matrix equations~\eqref{phieq},
i.e. find $f_{ij}, \chi_i$. Then the constants $k^i$ are given by
\ben\label{ki1}
k^i = \frac{1-x^2}{\det f} \, f^{ij} \partial_x \chi_j \, ,
\een
and this determines the full metric up to the choice of the remaining constant $C$.
We must furthermore ensure that, among all such solutions, we pick only those that give rise to a smooth
metric $g$.

The equations~\eqref{phieq} for $\Phi$ are easily integrated to
\ben
\Phi(x) = Q \, \exp \left[ 2 \, {\rm arcth} (x) \cdot L \right] = Q \, \left( \frac{1+x}{1-x} \right)^{L} \, \, .
\een
Here, $Q=\Phi(0), L= \frac{1}{2}(1-x^2) \Phi(x)^{-1} \partial_x \Phi(x)$ are both constant real $(D-2) \times (D-2)$ matrices, and we mean the matrix exponential etc.
It follows from the definition that $\Phi$
has the following general properties: It is symmetric, $\det \Phi = 1$, and it is positive definite. It is an easy consequence of these properties that $\det Q = 1$, ${\rm Tr} \, L = 0$ (taking the determinant of the equation), that
$Q = Q^T$ is positive definite, and that $L^T Q =  QL$. These relations allow us to write $Q = S^T S$ for some real invertible matrix $S = (s_{IJ})$ of determinant $\pm 1$, and to conclude that $SLS^{-1}$ is a real symmetric matrix. By changing $S$ to $VS$, where $V$ is a suitable orthogonal transformation, we can achieve that
\ben
SL S^{-1} = \left(
\begin{matrix}
\sigma_0 & 0 & \dots & 0\\
0 & \sigma_1 & \dots & 0\\
\vdots & & & \vdots\\
0 & 0 & \dots & \sigma_{D-3}
\end{matrix}
\right)
\een
is a real diagonal matrix, while leaving $Q$ unchanged. It then follows that
$\Phi(x) = S^T \, \exp \left[ 2 \, {\rm arcth} (x) \cdot S L S^{-1} \right] S$,
that is
\ben\label{phi}
\Phi_{IJ}(x) = \sum_{K=0}^{D-3} \left( \frac{1+x}{1-x} \right)^{\sigma_K} s_{KI} s_{KJ} \, .
\een
This is the most general solution to the field equation for $\Phi$ in the near horizon limit, and it depends
on the real parameters $s_{IJ}, \sigma_I$, which are subject to the constraints
\ben\label{constr}
\det (s_{IJ})
= \pm 1
\, , \quad \sum_{I=0}^{D-3} \sigma_I = 0 \, .
\een
The near horizon metric is completely fixed in terms of $\Phi$. It can be obtained combining eqs.~\eqref{phi} with eq.~\eqref{phidef} to determine $f_{ij}, \chi_i$, which in turn then fix the remaining constants $k^i, C$ in the near horizon metric. In the rest of this section, we explain how this can be done. It turns out that the smoothness of the near horizon metric also implies certain constraints on the parameters $\sigma_I, s_{IJ}$, and we will derive the form of these. Our analysis applies in principle to all dimensions $D \ge 4$. The case $D=4$, while being simplest, is somewhat different from the remaining
cases $D \ge 5$ and would require us to distinguish these cases in many of the formulae below. Therefore, to keep the discussion simple, we will stick to $D \ge 5$ in the following.

\medskip

First, we consider the $ij$-component of $\Phi$ in eq.~\eqref{phi}. By eq.~\eqref{phidef}
this is also equal to
\ben\label{phifconnect}
\sum_{I=0}^{D-3} \left( \frac{1+x}{1-x} \right)^{\sigma_I}
s_{Ii} s_{Ij} = \Phi_{ij} = f_{ij} + (\det f)^{-1} \, \chi_i \chi_j \, .
\een
Now, the coordinate $x \in [-1, 1]$
parametrizes the orbit space $H/U(1)^{D-3}$
of the horizon, which is topologically a finite interval. The boundary
points $x = \pm 1$ correspond to points on the horizon where an integer linear combination $\sum a^i_\pm \psi_i$ of the rotational Killing fields vanishes. 
This is equivalently expressed by the condition 
$f_{ij}(x) a_\pm^j \to 0$ as $x \to \pm 1$. By contrast, for all values of $x \in (-1,+1)$,
no linear combination of the rotational fields vanishes. Therefore, $\det f \neq 0$ for $x \in (-1,+1)$, while $\det f \to 0$ as $x \to \pm 1$. In fact, using eq.~\eqref{lambdax} one sees that
\ben
(\det f)^{-1} = 2 c_+^2(1-x)^{-1} + 2 c_-^2(1+x)^{-1} + \dots \quad
\text{as $x \to \pm 1$,}
\een
where the dots represent contributions that go to a finite limit, and where $c_{\pm}$ are non-zero constants related to $\lambda$ by $4c_\pm^2=\e^{-\lambda(\pm 1)} \neq 0$. The twist potentials $\chi_i$ also go to a finite limit as $x \to \pm 1$. By adding suitable constants to the twist potentials if necessary, we may achieve that
\ben
\chi_i \to \frac{1}{c_{\pm}} \, \mu_i \quad \text{as $x \to \pm 1$} \, ,
\een
where $\mu_i \in \mr$ are constants. The upshot of this discussion is that, as one approaches the boundary points, the components $\Phi_{ij}$ are dominated by the rank-1 part $(\det f)^{-1} \chi_i \chi_j$, which diverges as
$2(1 \mp x)^{-1} \, \mu_i \mu_j$ as $x \to \pm 1$. This behavior can be used to fix the possible values of the eigenvalues $\sigma_I$ as follows. First, it is clear that at least one of the eigenvalues must be non-zero, for otherwise
the right side of eq.~\eqref{phifconnect} would be smooth as $x \to \pm 1$, which we have just argued is not the case. Let us assume without loss of generality then that $\sigma_{D-3} \ge  \dots \ge \sigma_{D-3-n} > 0$ are the $n$ positive eigenvalues. Multiplying eq.~\eqref{phifconnect} by $1-x$ and taking $x \to +1$, we see that $\sigma_{D-3} = 1$, that $\mu_i = s_{(D-3)i}$, and that all other remaining positive eigenvalues must be strictly between 0 and 1. If we now subtract $(1-x^2)^{-1} \mu_i \mu_j$ from both sides of the equation, then the right side of eq.~\eqref{phifconnect} goes to a finite limit
as $x \to 1$, and so the left side has to have that behavior, too.
This is only possible if there are no other remaining positive eigenvalues besides $\sigma_{D-3}$. A similar argument then likewise shows that there is only one negative eigenvalue, which has to be equal to $-1$ (without loss of generality we may take $\sigma_{D-4} = -1$) and that $\mu_i = s_{(D-4)i}$.

In summary, we have shown that
\ben
\sigma_I = \begin{cases}
0 & \text{if $I\le D-5$,}\\
-1 & \text{if $I = D-4$,}\\
1 & \text{if $I = D-3$,}
\end{cases}
\een
and we also see that
\ben\label{Jeq}
\mu_i = s_{(D-3)i} = s_{(D-4)i} \, , \quad c_+ = s_{(D-3)0} \, , \quad c_- = s_{(D-4)0} \,.
\een
The condition that $\det S = \pm 1$ then moreover gives
\ben\label{Sdet}
\pm 1 = (c_+ - c_-) \, \epsilon^{ijk \dots m} s_{0i} s_{1j} s_{2k} \cdots \mu_m \,\, .
\een
We may now combine this information with the equations~\eqref{phi}
and~\eqref{phidef} and solve for $f_{ij},\chi_i$. The result can be expressed as:
\bena\label{fij}
f_{ij} \xi^i \xi^j &=& 2 \frac{1+x^2}{1-x^2} (\mu \cdot \xi)^2 + \sum_{I=0}^{D-5} (s_{I} \cdot \xi)^2 \\
&& - \frac{\e^{\lambda(x)}}{1-x^2} \left( (1-x^2) \sum_{I=0}^{D-5} \non
s_{I0} (s_{I} \cdot \xi) + [c_+(1+x)^2 + c_-(1-x)^2] (\mu \cdot \xi)
\right)^2\\
\chi_i \xi^i &=&\e^{\lambda(x)}\left( (1-x^2) \sum_{I=0}^{D-5}
s_{I0} (s_{I} \cdot \xi) + [c_+(1+x)^2 + c_-(1-x)^2] (\mu \cdot \xi)
\right) \, .
\eena
Here, we are using shorthand notations such as $\mu \cdot \xi = \mu_i \xi^i$ or
$s_I \cdot \xi = s_{Ii} \xi^i$, and
\ben\label{Fdef}
\exp[-\lambda(x)] = c_+^2(1+x)^2 + c_-^2(1-x)^2 +  (1-x^2) \sum_{I=0}^{D-5} s_{I0}^2 \, ,
\een
in order to have a reasonably compact notation. This function $\lambda$ agrees with
that previously defined in eq.~\eqref{lamdef} by eq.~\eqref{lambdax}.
From eq.~\eqref{fij}, one now finds after a short calculation that
the conditions~\eqref{bndycond} are equivalent to
\ben\label{s0I}
s_{I0} \, \mu_i a_+^i = c_{+} \, s_{Ii} a_+^i \, , \quad
s_{I0} \, \mu_i a_-^i = c_{-} \, s_{Ii} a_-^i \, , \quad \text{for $I=0, \dots, D-5$}.
\een
Either of these equations ``$\pm$'' can be used to solve for $s_{I0}$, because\footnote{
Indeed, let us assume that, say $\mu_i a^i_+ = 0$. Then, since $c_+ \neq 0$,
we know that also $s_{Ii} a_+^i = 0$. It then would follow that
$0=\epsilon^{ijk \dots m} s_{0i} s_{1j} s_{2k} \cdots \mu_m$, which however is
in contradiction with eq.~\eqref{Sdet}.
} $\mu_i a_\pm^i \neq 0$ for both ``$\pm$''. We will do this in the following.

As we have explained, the constants $k^i$ in the near horizon metric are given by~\eqref{ki1}. A longer calculation using eqs.~\eqref{fij},~\eqref{s0I},~\eqref{Sdet} and~\eqref{Fdef} reveals that
\ben\label{ki}
k^i = \frac{2c_+ c_-}{c_+ - c_-} \left( \frac{a_+^i}{\mu_j a_+^j} + \frac{a_-^i}{\mu_j a_-^j} \right) \, .
\een
To avoid conical singularities in the near horizon metric~\eqref{nhg}, we must furthermore have\footnote{
Here the constants $a^i_\pm \in \mz$ are normalized so that the greatest common divisor
of $a_+^i, i=1, \dots, D-3$ is equal to 1, and similarly for $a^i_-$.}
\ben\label{cdet1}
\frac{(1-x^2)^2}{\det f \cdot f_{ij} a^i_\pm a^j_\pm} \to  C^2 \quad \text{as $x \to \pm 1$,}
\een
and this determines $C$.
A longer calculation using eqs.~\eqref{fij},~\eqref{s0I}
shows that
\ben\label{cdet}
C = \frac{4c_+^2}{(c_+ - c_-) \mu_i a^i_+} = \frac{4c_-^2}{(c_+ - c_-) \mu_i a^i_-} \, .
\een
Thus, we have determined all quantities $C, k^i, f_{ij}$ in the near horizon metric~\eqref{nhg}.
We substitute these, and make the final coordinate change
\ben
x = \cos \theta \, , \quad 0 \le \theta \le \pi \, .
\een
Then, after performing some algebraic manipulations, we get the following
result, which summarizes our entire analysis so far:

\begin{thm}\label{thm1}
All non-static near horizon metrics (except topology type $H \cong T^{D-2}$)
are parametrized by the real parameters $c_\pm, \mu_i, s_{Ii}$,
and the integers $a_\pm^i$ where $I=0,\dots,D-5$ and $i=1, \dots, D-3$,
and ${\rm g.c.d.}(a_\pm^i) = 1$.
The explicit form of the near horizon metric in terms of these parameters is
\bena\label{NH}
g &=& \e^{-\lambda} (2\d v \d r - C^2 r^2 \d v^2 + C^{-2} \, \d \theta^2) +
\e^{+\lambda} \Bigg\{ (c_+-c_-)^2 (\sin^2 \theta) \, \Omega^2 \non\\
&& +(1+\cos \theta)^2 c_+^2 \sum_I \left( \omega_I - \frac{s_I \cdot a_+}{\mu \cdot a_+} \Omega \right)^2
+(1-\cos \theta)^2 c_-^2 \sum_I \left( \omega_I - \frac{s_I \cdot a_-}{\mu \cdot a_-} \Omega \right)^2\non\\
&& +  \frac{c_\pm^2\, \sin^2 \theta}{(\mu \cdot a_\pm)^2} \sum_{I < J} \Big(
(s_I \cdot a_\pm) \omega_J - (s_J \cdot a_\pm)\omega_I \Big)^2
\Bigg\} \, .
%
\eena
Here, the sums run over $I,J$ from $0, \dots, D-5$, the function
$\lambda(\theta)$ is given by
\ben
\exp[-\lambda(\theta)] = c_+^2(1+\cos \theta)^2 + c_-^2(1-\cos \theta)^2 + \frac{c_\pm^2 \sin^2 \theta}{(\mu\cdot a_\pm)^2} \sum_I (s_{I}\cdot a_{\pm})^2 \, ,
\een
$C$ is given by $C = 4c^2_\pm[(c_+-c_-)(\mu \cdot a_\pm)]^{-1}$, and we have defined  the 1-forms
\bena
\Omega(r) &=& \mu \cdot \d \varphi + 4Cr\frac{c_+c_-}{c_+ -c_-} \d v \\
\omega_I(r) &=& s_{I} \cdot \d \varphi +
\frac{r}{2} \, C^2 (s_{I} \cdot a_+ + s_I \cdot a_-)  \, \d v \, .
\eena
We are also using the shorthand notations such as $s_{Ii} a^i_+ = s_I \cdot a_+$, or
$\mu \cdot \d \varphi = \mu_i \d \varphi^i$, etc.
The parameters are subject to the constraints $\mu \cdot a_\pm \neq 0$ and
\ben\label{constr}
\frac{c_+^2}{\mu \cdot a_+} = \frac{c_-^2}{\mu \cdot a_-}
\, ,
\quad
\frac{c_+ (s_{I} \cdot a_+)}{\mu \cdot  a_+} = \frac{c_- (s_{I} \cdot a_-)}{ \mu \cdot  a_-}
\, ,
\quad
\pm 1 = (c_+ - c_-) \, \epsilon^{ijk \dots m} s_{0i} s_{1j} s_{2k} \cdots \mu_m
\een
but they are otherwise free. The coordinates $\varphi^i$ are $2\pi$-periodic,
$0 \le \theta \le \pi$, and $v,r$ are arbitrary. When writing ``$\pm$'', we mean that the
formulae hold for both signs.
\end{thm}
{\bf Remarks:}
(1) The function $\lambda(\theta)$ was invariantly defined in eq.~\eqref{lamdef}, and therefore
evidently has to be a smooth function. This is manifestly true, because both $c_\pm \neq 0$.
Because also $\mu \cdot a_\pm$ are both non-zero, we
explicitly see that the above metrics are smooth (in fact analytic).

(2) The part $2 \d v \d r - C^2 r^2 \d v^2$ of the metric is that of $AdS_2$ with curvature $C^2$. This is the cause for the
enhanced symmetry group of $O(2,1) \times U(1)^{D-3}$.

\medskip
\noindent
Let us finally discuss the meaning of the parameters on which the near horizon metrics depend.
The parameters $a_\pm^i \in \mz$ are related to the horizon topology.
Up to a globally defined coordinate transformation of the form
$$\varphi^i \mapsto \sum A^i_j \varphi^j \,\, {\rm mod} \,\, 2\pi \, , \quad A \in SL(\mz, D-3) \, ,$$
we have
\ben\label{apm}
a_+ = (1,0,0,\dots,0) \, , \quad a_- = (q, p, 0,\dots,0) \, , \quad p,q \in \mz \, , \quad {\rm g.c.d.}(p,q) = 1\, .
\een
A general analysis of compact manifolds with a cohomogeneity-1 torus action (see e.g.~\cite{hs}) implies that
the topology of $H$ is
\ben
H \cong
\begin{cases}
S^3 \times T^{D-5} & \text{if $p=\pm 1,q=0$,} \\
S^2 \times T^{D-4} & \text{if $p=0, q= 1$,}\\
L(p,q) \times T^{D-5} & \text{otherwise.}
\end{cases}
\een
The constants $\mu_i, c_\pm, a^i_\pm$ are directly related to the
horizon area by
\ben
A_H = \frac{(2\pi)^{D-3}(c_+-c_-)^2(\mu \cdot a_\pm)^2}{8c_\pm^4} \, ,
\een
and we also have
\ben
J_i := \frac{1}{2} \int_H \star (\d \psi_i) = (2\pi)^{D-3} \, \frac{c_+ - c_-}{2c_-c_+} \mu_i \, .
\een
In an asymptotically flat or Kaluza-Klein black hole spacetime with a single horizon $H$, the above integral for
$J_i$ could be converted to a convergent integral over a cross section at infinity using
Stokes theorem and the vanishing of the Ricci tensor. Then the $J_i$ would be equal to the
Komar expressions for the angular momentum. The near horizon limits that we consider do not
of course satisfy any such asymptotic conditions, and hence this cannot be done. Nevertheless,
if the near horizon metric under consideration arises from an asymptotically flat or
asymptotically Kaluza-Klein spacetime, then the $J_i$ are
the angular momenta of that spacetime. Hence, we see that the parameters $c_\pm, \mu_i, a_\pm^i$ are directly
related to geometrical/topological properties of the metric. This seems to be less clear for
the remaining parameters $s_{Ii}$.

The number of continuous parameters on which our metric depend can 
be counted as follows. 
First, the matrix $s_{Ii}$ has $(D-3)(D-4)$ independent components, 
$\mu_i$ has $(D-3)$ and $c_\pm$ has $2$ components. These parameters 
are subject to the $(D-2)$ constrains, eqs.~(\ref{constr}). 
However, changing $s_{Ii}$ to $\sum_{J=0}^{D-5} R^J{}_I s_{Ji}$, with 
$R^J{}_I$ an orthogonal matrix in $O(D-4)$, does not change the metric. 
Since such a matrix depends on $(D-4)(D-5)/2$ parameters, 
our metrics depend only on $(D-3)(D-4)+(D-3)+2 - (D-2) - (D-4)(D-5)/2
= (D-2)(D-3)/2$ real continuous parameters.

It is instructive to compare this number to the number of 
parameters of a boosted Kerr-brane. If we start from a direct 
product of a $4$-dimensional extremal Kerr metric with a flat torus 
$T^{D-4}$ and apply a boost in an arbitrary direction, 
then the resulting family of 
metrics has $(D-2)(D-3)/2$ 
parameters, and the horizon topology is $S^2 \times T^{D-4}$.  
It is plausible that all our metrics in our Thm.~\ref{thm1} for 
this topology can be obtained by taking the near horizon limit of 
these boosted Kerr-branes. 
By contrast, if we start with a direct product of a $5$-dimensional extremal 
Myers-Perry black hole with a flat torus 
$T^{D-5}$, then we similarly get a family of metrics 
which depends only on $(D-3)(D-4)/2 + 1$ parameters. 
Therefore in this case, we get metrics depending on fewer parameters 
than those in Thm.~\ref{thm1}. 

\section{Examples}

Let us first illustrate our classification in $D=5$ spacetime dimensions. According to our general result,
the metrics have the discrete parameters $a_\pm^1, a_\pm^2$ as well as the 6
continuous parameters $\mu_1, \mu_2, s_{01}, s_{02}, c_+, c_-$ which are subject to 3 constraints.
Thus, the number of free parameters is 3, and we take $C$ [given by eq.~\eqref{cdet}] as one of them
for convenience. We have the following cases to consider,
depending on the possible values of the discrete parameters, see eq.~\eqref{apm}:

\medskip
\noindent
{\bf Topology $H \cong S^1 \times S^2$}: This case corresponds to the choice $a_+ = a_- = (1,0)$.
The constraints~\eqref{constr} read explicitly
\ben
c_+^2 \mu_1 = c_-^2 \mu_1 \, , \quad
c_+ s_{01} \mu_1 = c_- s_{01} \mu_1 \, , \quad
(c_+ - c_-)
\left|
\begin{matrix}
\mu_1 & s_{01}\\
\mu_2 & s_{02}
\end{matrix}
\right|
= 1 \,
\een
in this case.
We know that $\mu_1$ cannot vanish, so the first and third
equation imply together that $c_\pm = \pm B$ for some non-zero constant $B$. As a consequence,
the second equation then gives $s_{01} = 0$, from which the third equation then gives
$s_{02} = 1/(2c_+\mu_1)$. Putting all this into our formula~\eqref{NH} for the near horizon metric
gives
\bena
g &=& 2B^2(1+\cos^2 \theta)(2\d v \d r - C^2 r^2 \d v^2 + C^{-2} \d \theta^2) +
\frac{C^2}{16 B^4} (\d \varphi^2)^2 \non\\
&&+ \frac{8B^2\sin^2 \theta}{C^2(1+\cos^2 \theta)} \left(\d \varphi^1 + A \, \d \varphi^2 +  C^2 r \, \d v\right)^2 \, ,
\eena
where we have put $A = \mu_2/\mu_1$.
We can explicitly read off from the metric that the norm of
$\partial/\partial \varphi^1$ [i.e., the coefficient of $(\d \varphi^1)^2$] vanishes at $\theta=0,\pi$, whereas the norm of
$\partial/\partial \varphi^2$ [i.e., the coefficient of $(\d \varphi^2)^2$]
never vanishes. This is the characteristic feature of the action of $U(1)^2$ on $S^2 \times S^1$.

\medskip
\noindent
{\bf Topology $H \cong S^3$}: In this case, $a_+ = (1,0), a_-=(0,1)$.
The constraints~\eqref{constr} are
\ben
c_+^2 \mu_2 = c_-^2 \mu_1 \, , \quad
c_+ s_{01} \, \mu_2 = c_- s_{02} \, \mu_1 \, , \quad
(c_+ - c_-)
\left|
\begin{matrix}
\mu_1 & s_{01}\\
\mu_2 & s_{02}
\end{matrix}
\right|
= 1 \, .
\een
The constraints allow us e.g. to express $\mu_1, \mu_2, s_{01}, s_{02}$ in terms of
$A:=c_+, B:=c_-$ and $C$ given by eq.~\eqref{cdet}. The result must then be plugged back into
the equation for the near horizon metric~\eqref{NH}. After some calculation,
one ends up with the result
\bena\label{gs3}
g &=& \e^{-\lambda} (2\d v \d r - C^2 r^2 \d v^2+ C^{-2} \, \d \theta^2 ) \\
&&+ \e^{+\lambda} \Bigg\{
\left(\frac{4}{C}\right)^2 \sin^2 \theta \left( A^2 \d \varphi^1 + B^2 \d \varphi^2 + r ABC^2 \d v \right)^2 \non\\
&&+
\left(\frac{C}{4}\right)^2 (1+\cos \theta)^2 \left( A^{-1} \, \d \varphi^2  + r(2B)^{-1} C^2  \, \d v\right)^2\non\\
&&+
\left(\frac{C}{4}\right)^2 (1-\cos \theta)^2 \left( B^{-1} \, \d \varphi^1 +  r(2A)^{-1} C^2  \, \d v\right)^2
\Bigg\} \, , \non
\eena
where
\bena
\exp[-\lambda(\theta)] &=& A^2(1+\cos \theta)^2 + B^2(1-\cos \theta)^2 + \left( \frac{C^2}{16 AB} \right)^2 \sin^2 \theta \, .
\eena
The quantity $A-B$ must be non-zero on account of the third constraint.
Note that $\exp \lambda(\theta) \neq 0$ for $0 \le \theta \le \pi$, so we can explicitly read off from the metric that
the norm of $\partial/\partial \varphi^2$ [i.e., the coefficient of $(\d \varphi^2)^2$] vanishes at
$\theta = \pi$, whereas the norm of $\partial/\partial \varphi^1$ [i.e., the coefficient of $(\d \varphi^1)^2$] vanishes at
$\theta = 0$. This is the characteristic feature of the action of $U(1)^2$ on the 3-sphere.



\medskip
\noindent
{\bf Topology $H \cong L(p,q)$}: In this case, $a_+ = (1,0), a_-=(q,p)$, where $p,q \in \mathbb Z$ and $p \neq 0$.
The constraints~\eqref{constr} are explicitly
\ben
c_+^2 (q\mu_1+p\mu_2) = c_-^2 \mu_1 \, , \quad
c_+ s_{01} \, (q\mu_1+p\mu_2) = c_- (qs_{01} + ps_{02}) \, \mu_1 \, , \quad
(c_+ - c_-)
\left|
\begin{matrix}
\mu_1 & s_{01}\\
\mu_2 & s_{02}
\end{matrix}
\right|
 = 1 \, .
\een
We choose as the independent parameters $A:=  c_+/p, B:= c_-/p$, and 
$C$ given by eq.~\eqref{cdet}, and solve for the remaining ones 
using the constraints. The result is plugged back into 
the equation for the near horizon metric~\eqref{NH}. After some calculation, 
one ends up with the result 
\bena\label{glpq}
g &=& \e^{-\lambda}(2\d v \d r - C^2 r^2 \d v^2 + C^{-2} \d \theta^2)
\\
&&+ p^2 \e^{+\lambda}\Bigg\{
\left(\frac{4p}{C}\right)^2 \sin^2 \theta \left( A^2 (1/p) \d \varphi^1 + B^2 (\d \varphi^2 - (q/p)\d \varphi^1) + rABC^2 \d v \right)^2 \non\\
&&+
\left(\frac{C}{4p}\right)^2 (1+\cos \theta)^2 \left( A^{-1}(\d \varphi^2 - (q/p) \d \varphi^1) + r(2B)^{-1} C^2 \, \d v\right)^2\non\\
&&+
\left(\frac{C}{4p}\right)^2 (1-\cos \theta)^2 \left( (pB)^{-1} \d \varphi^1
+r(2A)^{-1} C^2  \, \d v\right)^2
\Bigg\}\non \, ,
\eena
where
\bena
\exp[-\lambda(\theta)] &=&
p^2\Big[
A^2(1+\cos \theta)^2 + B^2(1-\cos \theta)^2 + \left( \frac{C^2}{16p^2 AB} \right)^2 \sin^2 \theta
\Big]\, .
\eena
We note that at $\theta = \pi$, the Killing field $\partial/\partial \varphi^1$ has vanishing norm,
while at $\theta = 0$, the Killing field $q \partial/\partial \varphi^1 + p \partial/\partial \varphi^2$ has vanishing norm. This is the characteristic feature of the action of $U(1)^2$ on the Lens space $L(p,q)$.

The metrics with $H \cong L(p,q)$ just described are closely related to those in the case $H \cong S^3$ described in
the previous example. Indeed, in the case $H \cong S^3$,
consider the map given by $(\varphi^1, \varphi^2) \mapsto (\varphi^1 + 2\pi/p, \varphi^2 + 2\pi q/p)$,
leaving invariant the other coordinates, where $\varphi^1, \varphi^2$ are $2\pi$-periodic. This map is an isometry of the metric with $H \cong S^3$, and by repeated application
generates the subgroup $\mz_p$ of the full isometry group. If we factor by this group,
then we get a metric with $H \cong L(p,q)$, and we claim that this metric is exactly the one just given.
To see this more explicitly, we note that factoring by the above group $\mz_p$ of isometries in effect
imposes the further identifications
\ben\label{ident}
(\varphi^1, \varphi^2) \cong (\varphi^1 + 2\pi/p, \varphi^2 + 2\pi q/p)
\een
on the angular coordinates in the metric~\eqref{gs3}, which were initially $2\pi$-periodic. If we let
\ben
f: (r,v,\theta,\varphi^1, \varphi^2) \mapsto (r, p^2 v, \theta, (1/p) \varphi^1, \varphi^2 - (q/p) \varphi^1) \,
\een
then $f$ provides an invertible mapping from the ordinary $2\pi$-periodic coordinates to the coordinates with the identifications~\eqref{ident}.
If we now take the metric~\eqref{gs3} in the case $H \cong S^3$, factor it by
$\mz_p$, pull it back by $f$, and furthermore put $C \to C/p$, then we get precisely the $H \cong L(p,q)$
metrics~\eqref{glpq}. Thus, all metrics in the case $H \cong L(p,q)$ arise from the case
$H \cong S^3$ by taking quotients. The same statement (with similar proof) is true in all dimensions $D$.


\vspace{1cm}

Let us finally briefly discuss an example of our classification in $D=6$ dimensions. In this case, the metrics
are classified by the discrete parameters $a_\pm$ [see eq.~\eqref{apm}] and $7$ real continuous parameters. An example is

\medskip
\noindent
{\bf Topology $S^3 \times S^1$:} In this case, $a_+ = (1,0,0), a_-=(0,1,0)$.  The constraints are explicitly
\ben
c_+ s_{01} \mu_2 = c_- s_{02} \mu_1 \, \quad
c_+ s_{11} \mu_2 = c_- s_{12} \mu_1 \, \quad c_+^2 \mu_2 = c_-^2 \mu_1 \, , \quad
(c_+ - c_-) \left|
\begin{matrix}
\mu_1 & s_{01} & s_{11}\\
\mu_2 & s_{02} & s_{12}\\
\mu_3 & s_{03} & s_{13}
\end{matrix}
\right| = 1 \, .
\een
To simplify the formulae somewhat, we consider the special case that $c_+ = -c_-=: A/2$.
Then the constraints may be solved easily for the remaining parameters. To obtain
a halfway simple expression, we also consider the special case $s_{11} = s_{03} = 0$,
and we denote the remaining free parameters as
$B:=s_{01}, D = \mu_3$, and $C$ as usual. The resulting metric is still rather
complicated and is given by
\bena\label{NH1}
g &=& \e^{-\lambda(\theta)} \left( 2\d v \d r - C^2 r^2 \d v^2 + C^{-2} \d \theta^2 \right) \non\\
%
&&+ \e^{+\lambda(\theta)} \Bigg\{
A^4 C^{-2} \sin^2 \theta \left( \d \varphi^1 + \d \varphi^2 + A^{-1}CD \, \d\varphi^3 - rC^2 \, \d v \right)^2 \non\\
&&+ \frac{A^2B^2}{4} (1+ \cos \theta)^2 (2 \d \varphi^2 + A^{-1} CD \, \d \varphi^3 - rC^2 \, \d v)^2 \non\\
&&+ \frac{A^2B^2}{4} (1- \cos \theta)^2 (2 \d \varphi^1 + A^{-1} CD \, \d \varphi^3 - rC^2 \, \d v)^2 
\Bigg\} 
+ \frac{C^2}{4A^4B^2} (\d \varphi^3)^2 \,.
\eena
Here we also have
\bena
\e^{-\lambda(\theta)} &=& \frac{A^2}{2} (1 + \cos^2 \theta)
+ \frac{B^2 C^2}{4} \, \sin^2 \theta \, .
\eena
This special family of metrics depends on only 4 parameters. 
It is easy to write down the general 7 parameter family of metrics.

\section{Conclusion}

We have determined explicitly what are the possible (non-static) stationary smooth, cohomogeneity-one
near horizon geometries satisfying the vacuum Einstein equations. We
 excluded by hand\footnote{ 
See, however, the note added in proof. 
} 
the case that the horizon topology is $T^{D-2}$.
The solution, described in thm.~\ref{thm1}, is given in closed form in
terms of real and discrete parameters (corresponding to the possible topology types other than $T^{D-2}$), which are subject to certain constraints that take
the form of algebraic equations. After taking into account these 
constraints, the metrics depend on
$(D-2)(D-3)/2$ independent real parameters, and two discrete ones. 
%
For example, in $D=5$, we initially have 3 real continuous parameters.
We have worked out explicitly this case as did~\cite{kl},
but our metrics are presented in different coordinates\footnote{We also do not distinguish between
the subcases ``A'' and ``B'' as in~\cite{kl} but instead give a unified expression for the metric.}
for the case $H \cong S^3$. In $D \ge 6$, not all of our metrics 
can be obtained as the near horizon limit of a known black hole 
solution, so in this sense some of our metrics are new for $D \geq 6$.

By contrast to $D \le 5$, not all near horizon metrics that we have found can be obtained as the near horizon limits of known black hole
solutions in dimensions $D \ge 6$. It is conceivable that there are further extremal black hole solutions---to be found---which give our metrics in the near horizon limit, but it is also possible that some of our metrics in
$D \ge 6$ simply do not arise in this way.

Our method as described only works for vacuum solutions. However, we expect that it can be generalized
to any theory whose equations can be recast into equations of the sigma-model type that we encounter.
Thus we expect our method to be applicable e.g. to 5-dimensional minimal supergravity, see e.g.~\cite{bou,virmani,Clement,tom}. By contrast,
our method does not seem applicable straightforwardly to the case of a cosmological constant.
In our proof, we also assumed that the metrics are {\em not} static. All static near horizon geometries were
found in~\cite{kl1} in $D=5$ and in \cite{crt06} in arbitrary dimensions. 


It would be interesting to see whether our classification can be used to prove a black hole uniqueness
theorem in arbitrary dimensions
for extremal black holes along the lines of~\cite{don,fl}, thereby generalizing~\cite{hs,hs1}. It would also
be interesting to investigate whether our analysis can be used to obtain new structural insights into the origin of the Bekenstein-Hawking
entropy, e.g. by considering a suitably quantized version of eq.~\eqref{phieq}.

\vspace{1cm}

{\bf Acknowledgements:} 
S.H. would like to thank the {\em Centro de Ciencias de Benasque Pedro Pascual}
for its hospitality during the inspiring programme on ``Gravity - New perspectives from strings and higher dimensions'',
where a key part of this work was done. He would also like to thank P. Figueras, H. Kunduri and especially
J. Lucietti for numerous useful discussions. 
We especially would like to thank the unknown referee for pointing out 
an error in the counting of parameters of our solutions and for 
suggesting a simplification of formula~(\ref{NH1}).  
This work is supported in part by the Grant-in-Aid for Scientific 
Research from the Ministry of Education, Science and Culture of Japan.

\vspace{2cm}

{\bf Note added in proof:} 
In our analysis, we excluded by hand the horizon topology 
$T^{D-2}$. There cannot exist any asymptotically flat or Kaluza-Klein
black hole solutions with this topology by general 
arguments~\cite{galoway, hs}. 
At any rate, these could not arise as the near horizon limits of 
a black hole. 
After we finished this work, it was confirmed by J. Holland that there 
cannot be any {\em non-static} cohomogeneity-one near horizon 
geometries with topology $H \cong T^{D-2}$ \cite{holland}. Hence our main 
theorem~\ref{thm1} covers {\em all} possibilities with $D-3$ commuting 
rotational symmetries. 
The static case is covered by the results of \cite{crt06}.

\end{document}